\documentclass[twocolumn]{aastex631}
\newcommand{\LyA}{Ly$\alpha$\,}
\newcommand{\lya}{\LyA}
\def\la{\mathrel{\hbox{\rlap{\hbox{\lower3.6pt\hbox{$\sim$}}}\hbox{\raise1.4pt\hbox{$<$}}}}}
\def\ga{\mathrel{\hbox{\rlap{\hbox{\lower3.6pt\hbox{$\sim$}}}\hbox{\raise1.4pt\hbox{$>$}}}}}
\def\ergcm2s{\ifmmode {\rm\,erg\,cm^{-2}\,s^{-1}}\else
                ${\rm\,ergs\,cm^{-2}\,s^{-1}}$\fi}
\def\ergsec{\ifmmode {\rm\,erg\,s^{-1}}\else
                ${\rm\,ergs\,s^{-1}}$\fi}

\def\Msun{M_\odot}

\newcommand{\hii}{H{\sc i}~21\,cm}
\newcommand{\hi}{H{\sc i}}

\newcommand{\cm}{cm$^{-2}$}
\newcommand{\kms}{km~s$^{-1}$}

\shorttitle{A Merger-Triggered Starburst in a Green Pea Galaxy}
\shortauthors{Purkayastha et al.}

\begin{document}

\title{The Second Case of a Major Merger Triggering a Starburst in a Green Pea Galaxy}

\author{S. Purkayastha}
\affiliation{National Centre for Radio Astrophysics, Tata Institute of Fundamental Research, Pune University, Pune 411007, India}

\author{N. Kanekar}
\affiliation{National Centre for Radio Astrophysics, Tata Institute of Fundamental Research, Pune University, Pune 411007, India}

\author{S. Kumari}
\affiliation{National Centre for Radio Astrophysics, Tata Institute of Fundamental Research, Pune University, Pune 411007, India}

\author{J. Rhoads}
\affiliation{Astrophysics Division, NASA Goddard Space Flight Center, Greenbelt, MD 20771, USA}

\author{S. Malhotra}
\affiliation{Astrophysics Division, NASA Goddard Space Flight Center, Greenbelt, MD 20771, USA}

\author{J. Pharo}
\affiliation{Leibniz Institute for Astrophysics Potsdam (AIP), An der Sternwarte 16,
14482 Potsdam, Germany}

\author{T. Ghosh}
\affiliation{Green Bank Observatory, P.O. Box 2, Green Bank, WV 24944, USA}

\begin{abstract}
	We have used the Karl G. Jansky Very Large Array (VLA) to map H{\sc i}~21\,cm emission from the Green Pea galaxy GP~J1148+2546 at $z \approx 0.0451$, only the second measurement of the H{\sc i} spatial distribution of a Green Pea. The VLA H{\sc i}~21\,cm image, the DECaLS optical image, and Sloan Digital Sky Survey spectroscopy show that GP~J1148+2546 has two neighbours, the nearer of which is only $\approx17.5$~kpc away, and that the H{\sc i}~21\,cm emission extends in an inverted ``C'' shape around the Green Pea and its companions, with the highest H{\sc i} column density between the two neighbouring galaxies. The starburst in GP~J1148+2546 is likely to have been triggered by the ongoing merger with its neighbours, although the velocity field and velocity dispersion images do not show clear merger signatures at the Green Pea location. The H{\sc i} mass of the Green~Pea and its immediate surroundings is $(3.58\pm0.37)\times{10^9}\,M_\odot$, a factor of $\approx7.4$ lower than the total H{\sc i} mass of the system of three interacting galaxies, while the H{\sc i} depletion timescale of GP~J1148+2546 is $\approx0.69$~Gyr, much shorter than that of typical galaxies at $z\approx0$. We detect damped Ly$\alpha$ absorption and Ly$\alpha$ emission from the Green Pea in a Hubble Space Telescope Cosmic Origins Spectrograph spectrum, obtaining a high H{\sc i} column density, $\approx2.0\times10^{21}$~cm$^{-2}$, and a low Ly$\alpha$ escape fraction, $\approx0.8$\%, consistent with the relatively low value ($\approx5.4$) of the ratio O32~$\equiv$~[O{\sc iii}]$\lambda 5007 + \lambda 4959$/[O{\sc ii}]$\lambda$3727,3729.

\end{abstract}


\section{Introduction} \label{sec:intro}

Understanding the mechanisms responsible for the last unexplored observable phase of cosmic evolution, the Epoch of Reionization (EoR), has been a long-standing quest in astronomy. During this epoch, $z \gtrsim 6$, the mostly neutral intergalactic medium (IGM) is thought to have been re-ionized primarily by Lyman-continuum (LyC) photons leaking from faint, star-forming, dwarf galaxies \citep[e.g.][]{fan2006observational}. While direct observations of these high-redshift dwarf galaxies are difficult due to limitations in both sensitivity and resolution, nearby analogs with similar properties provide an exciting proxy to understand the processes that led to cosmic reionization. Over the last decade, efforts have hence been made to identify local analogs of EoR galaxies, 
which frequently show strong \lya\ emission indicative of extreme star formation properties \citep{malhotra2002ew}, accompanied by low metal abundances \citep{maseda2023metals}. 
Green Pea galaxies (GPs), discovered by citizen scientists in the Galaxy Zoo project \citep{cardamone2009galaxy} using the Sloan Digital Sky Survey (SDSS), are today considered to be the best local analogs of the high-$z$ dwarf galaxies that caused cosmic reionization.

GPs are extreme emission-line dwarf starburst galaxies, with low metallicity, low dust content, high specific star-formation rate (sSFR), compact size, and high ionization, found at redshifts of $z \lesssim 0.3$ \citep[e.g.][]{cardamone2009galaxy,amorin2010oxygen,nakajima2014ionization,jiang2019gp,kim2020sfi,kim2021gpsize}. Most GPs have their Ly$\alpha$ line in emission \citep[e.g.][]{henry2015lyalpha,yang2016green,yang2017lyalpha,jaskot2019,izotov2020}, with a  Ly$\alpha$ rest equivalent width distribution similar to that of high-$z$ Ly$\alpha$ emitters \citep[e.g.][]{yang2017lyalpha}. Even more interestingly, GPs have been found to show LyC leakage, with leakage fractions of $\approx 2-72$\% \citep[e.g.][]{izotov2016,izotov2016detection,izotov2018j1154+,izotov2018low}. These properties are similar to those of the early galaxies that reionized the Universe \citep[e.g.][]{rhoads23gphiz}, and that form the basis for \lya-based tests of reionization \citep{malhotra2004reion}.

One of the perplexing features of GPs is the 
combination of high sSFR with LyC and \lya\ leakage.
This raises tension between the need for cold neutral gas to fuel the star formation and the requirement for a sufficiently low  \hi\ column density to allow the LyC and resonantly-scattered \lya\  emission to escape. Mapping the \hi\ distribution is crucial to understand the physics of GPs and the escape of the \lya\ and LyC photons.

Recently, \cite{kanekar2021atomic} used the Arecibo Telescope and the Green Bank Telescope to obtain the first detections of \hii\ emission from 19 Green Peas  at $z \approx 0.02 - 0.1$, opening the possibility of \hii\ mapping studies to probe the \hi\ distribution in GPs. Following this, \citet{purkayastha2022} used the Giant Metrewave Radio Telescope (GMRT) to carry out the first \hii\ mapping of a Green Pea (GP~J0213+0056, at $z \approx 0.0399$), finding clear evidence of a major merger interaction between the GP and a companion galaxy ($\approx 4.7$~kpc away), that is likely to have triggered the GP starburst and also driven out the \hi\ from the GP, resulting in the leakage of \lya\ photons.

In this paper, we present Karl G. Jansky Very Large Array (VLA) \hii\ mapping of a second Green Pea, GP~J1148+2546 at $z=0.0451$. This galaxy has the highest \hi\ mass of the GP sample of \citet{kanekar2021atomic}, and is an outlier in their $\rm M_{HI} - M_{B}$ relation. We also present Hubble Space Telescope (HST) Cosmic Origins Spectrograph (COS) \lya\ spectroscopy of the GP, aiming to connect the \lya\ and optical properties of the GP to its \hi\ distribution. {Throughout this paper, we will asssume a Planck-2018 $\Lambda$ cold dark matter cosmology, with $H_0 = 67.4$~\kms~Mpc$^{-1}$, $\Omega_m = 0.315$ and $\Omega_\Lambda = 0.685$ \citep{planck20}. For this cosmology, the luminosity distance of GP~J1148+2546 is 207.398~Mpc.}

\begin{figure*}
    \centering
    \includegraphics[width=0.475\textwidth]{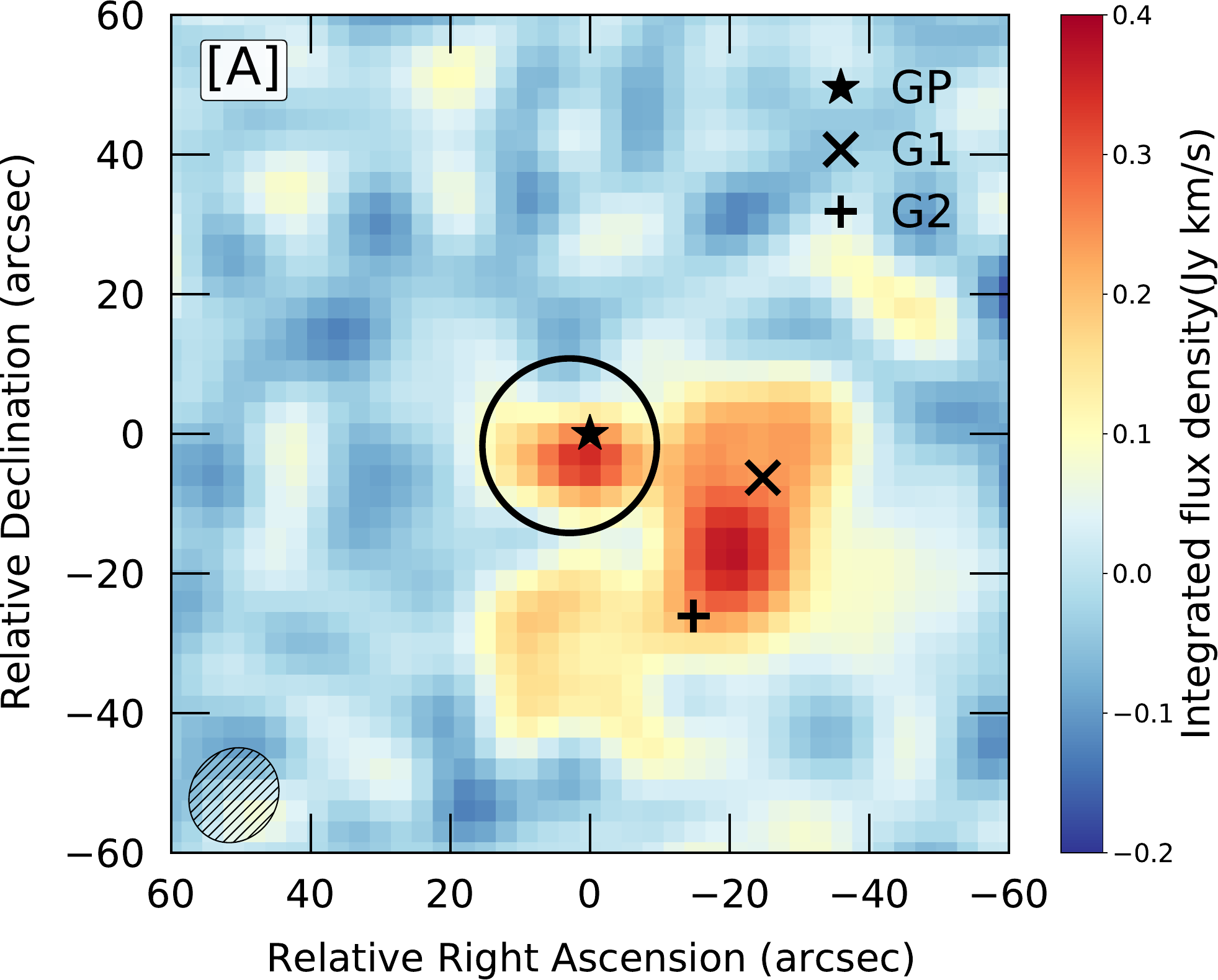}
    \includegraphics[width=0.475\textwidth]{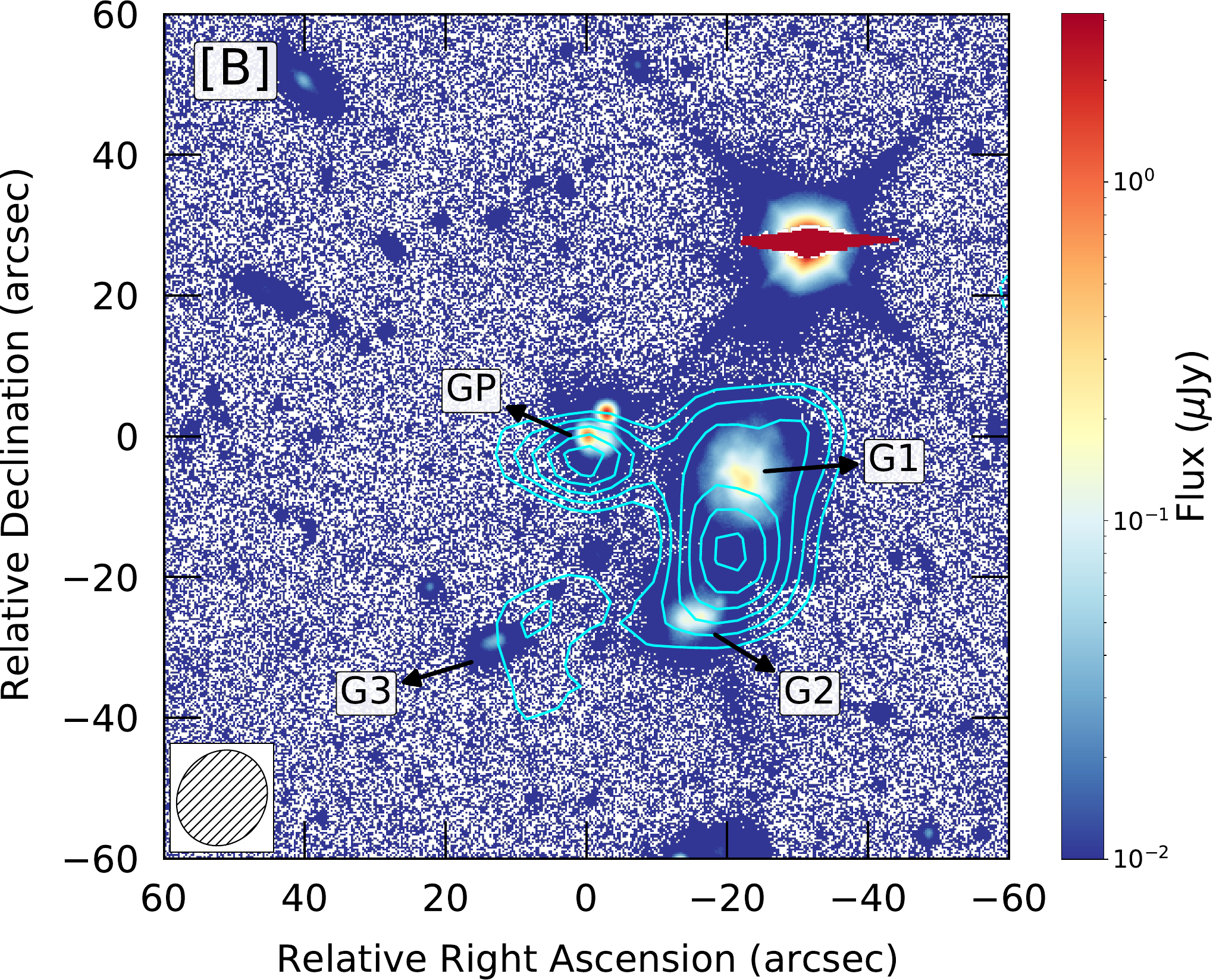}
    \caption{[A]~(left panel) VLA \hi\ intensity image of the field of GP~J1148+2546, at a resolution of $14\arcsec~\times~12\arcsec$. The locations of GP~J1148+2546 and its two companions are indicated by the star, the ``X'', and the ``+'' symbols, respectively.
    [B]~(right panel) The VLA \hi\ intensity image (in contours) overlaid on the DECaLS g-band image (in colour). The contour levels are [-3.0,3.0,4.0,5.0,6.0,7.0,8.0]~$\times \sigma$, {where $\sigma = 0.045$~Jy~Beam$^{-1}$~km~s$^{-1}$ is the RMS noise on the image, corresponding to an \hi\ column density of $3.1\times 10^{20}$~\cm}. The locations of GP~J1148+2546 and its companions are indicated by the labels GP, G1, and G2 {(and the possible third companion by G3)}. In both panels, the coordinates are relative to the J2000 coordinates of GP~J1148+2546.}
    \label{fig:gp1148-mom0}
\end{figure*}

\section{Observations and Data Analysis} \label{sec:obs}

\subsection{VLA \hii\ spectroscopy}

We observed GP~J1148+2546 (J2000 co-ordinates: 11h48m27.3s, +25d46'11.7'') with the VLA L-Band receivers on 11 March 2020 (proposal VLA/20A-242; PI Kanekar), in the C-array configuration. The observations used a bandwidth of 32 MHz, centred at 1358.9~MHz and sub-divided into 2048 channels, with two circular polarizations. The total on-source time was $\approx 6.6$~hours. Observations of 3C286 and 3C147 were used to calibrate the flux density scale, and of J1156+3128 to measure the antenna-based gains and the antenna bandpass shapes.

The data were analysed using standard procedures in the Common Astronomy Software Applications {\sc casa} package \citep[version 5.6;][]{mcmullin2007casa}. We initially excluded all visibilities from non-working antennas or those affected by radio frequency interference (RFI) from the data set. We then used the calibrator data to determine the antenna-based gains and bandpass shapes, using the routines {\sc gaincalR} and {\sc bandpassR} \citep{chowdhury20}, and applied the gains and bandpass solutions to the target data. We then made a continuum image of the field of GP~J1148+2546 from the line-free channels, using the routine {\sc tclean}, the w-projection algorithm, and Briggs weighting \citep[with robust$=-0.5$;][]{briggs95}. A standard iterative self-calibration procedure was then used, with 6 rounds of phase-only self-calibration and imaging, and 3 rounds of amplitude-and-phase self-calibration and imaging, along with inspection of the residual visibilities and further data editing to remove any data affected by RFI. The procedure was carried out until no improvement was seen in either the image or the residual visibilities on further self-calibration. The final continuum image has an angular resolution (full width at half maximum, FWHM, of the synthesized beam) of $14\arcsec \times 12\arcsec$ and a root-mean-square (RMS) noise of $\approx 36 \ \mu$Jy~Beam$^{-1}$. {After applying the final gain solutions to the multi-channel visibilities, we used the routine {\sc uvsub} to subtract the cleaned continuum emission from the self-calibrated visibilities, to obtain a residual visibility data set.}

The next step in the analysis is to make spectral cubes with the {\sc tclean} routine, using different weighting schemes, trading between sensitivity and resolution. Before doing this, for each resolution, we made a continuum image from the residual visibilities at the same resolution, again using the routine {\sc uvsub}, and subtracted it out, to remove any residual continuum emission. The cubes were made in the barycentric frame, with a velocity resolution of 10~\kms, and were cleaned to 0.5 times the per-channel RMS noise.  

We first made a cube using natural weighting, where the fatter beam ($21\arcsec \times 17\arcsec$) and higher sensitivity allows us to pick up more emission, yielding a more accurate estimate of the total \hi\ mass. We made a second cube with uniform weighting, where the narrow synthesized beam ($14\arcsec \times 12\arcsec$) yields a higher-resolution image of the \hi\ distribution. For each cube, we finally used the task {\sc imcontsub}, to fit, and subtract out, a linear baseline to line-free channels at each spatial pixel of the cubes. {The RMS noise on the final naturally-weighted cube is 0.42~mJy~Beam$^{-1}$ per 10~\kms\ channel, while that on the final uniformly-weighted cube is $\approx 0.76$~mJy~Beam$^{-1}$ per 10~\kms\ channel.} 

{The velocity range $-174$~\kms\ to $+136$~\kms\ was used to make the moment images and calculate the total velocity-integrated \hii\ line flux density. This range was chosen based on the full width between nulls of the Arecibo telescope single-dish \hii\ spectrum \citep{kanekar2021atomic}. The channel maps for each 10~\kms\ channel in this velocity range are shown in Fig.~\ref{fig:channelmap} in the appendix.}

{We used the routine {\sc immoments} in {\sc casa} to obtain the zeroth velocity moment, i.e. the velocity-integrated \hii\ line flux density, of each spectral cube, to study the \hi\ spatial distribution. No blanking of pixels was carried out while making the zeroth moment image. To examine the kinematic properties, the first and second moment images (i.e. the \hii\ intensity-weighted velocity field and the velocity dispersion, respectively) were made in the package {\sc aips} \citep{greisen03}, using the task {\sc momnt}. These images were made from the same velocity channels, but after blanking pixels with emission having $\lesssim 2\sigma$ significance after smoothing the cube by 3 channels. }

\subsection{HST COS UV Spectroscopy}

We used archival spectra from the HST COS instrument\footnote{The HST data used in this paper can be found in MAST: \dataset[10.17909/0khz-tg92]{http://dx.doi.org/10.17909/0khz-tg92}.} to obtain UV coverage of the redshifted Ly$\alpha$ line line from GP~J1148+2546. The COS Legacy Archive Spectroscopic SurveY \citep[CLASSY, HST prog.~15840;][]{berg2022cos} observed GP~J1148+2546 for 4552.32s with the G130M FUV grating at a central wavelength of 1222~\AA. This mode has a resolving power of $R>10000$ at the observed Ly$\alpha$ wavelength $\lambda=1271$~\AA\ and a spectral dispersion of $\sim 10\, \hbox{m\AA}\,\hbox{pixel}^{-1}$. The Segment~A spectrum has a wavelength range of $1220$~\AA~$ < \lambda < 1360$~\AA. After checking for and finding no listed data quality issues for the observations in the CLASSY archive in MAST, we extracted the spectrum from the \textit{x1dsum} file produced by the standard CalCOS pipeline. As the spectrum shows Ly$\alpha$ both in emission and absorption, we first fit a skewed Gaussian to the emission component using the \textit{lmfit} package \citep{newville2015} with an initial guess centered on the line peak. The best-fit emission line is then subtracted from the spectrum, leaving the absorption component, which we fit with a Voigt profile to measure the \hi\ column density.

\section{Results}
\label{sec:results}

\begin{figure*}
    \centering
    \includegraphics[width=0.495\textwidth]{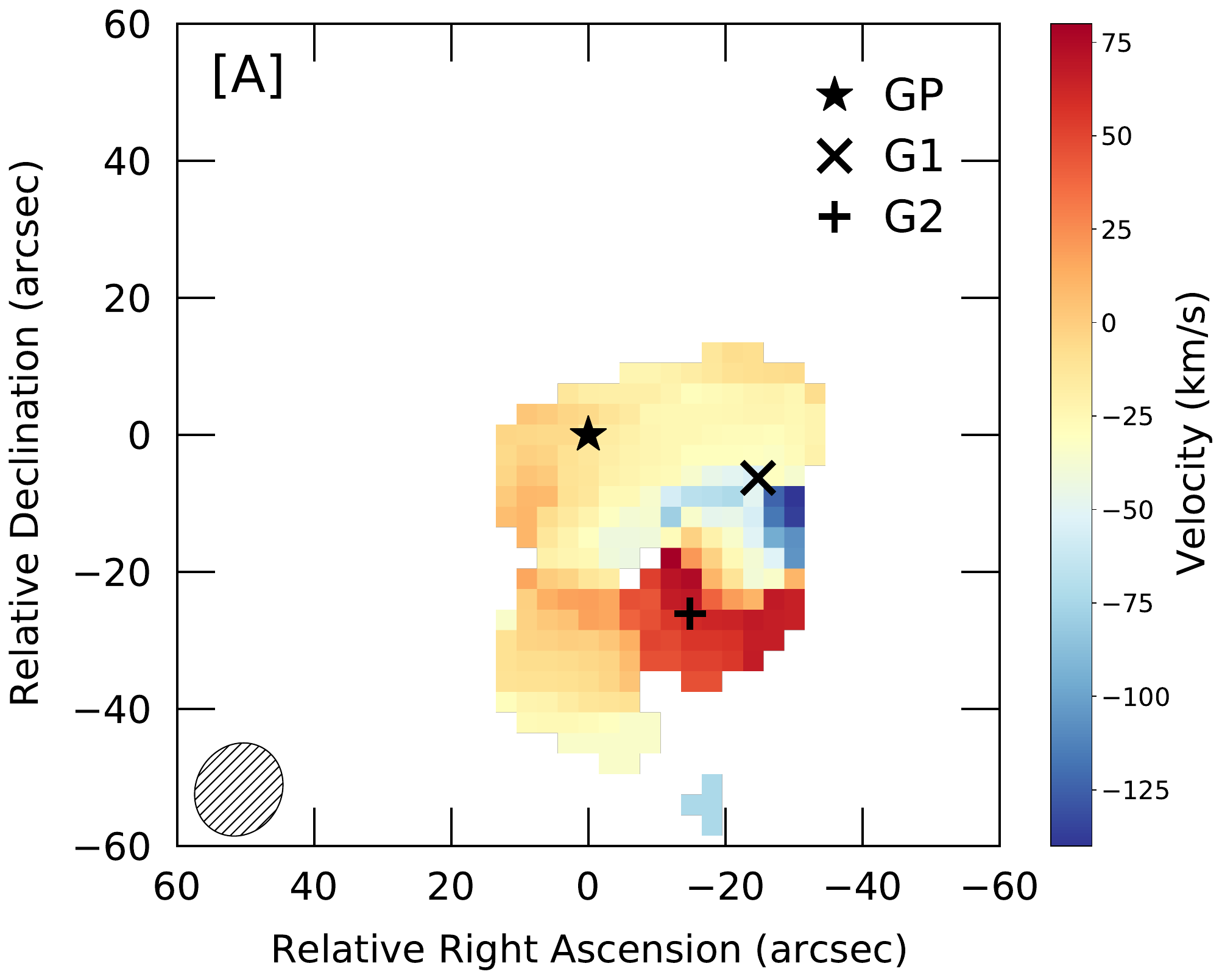}
    \includegraphics[width=0.495\textwidth]{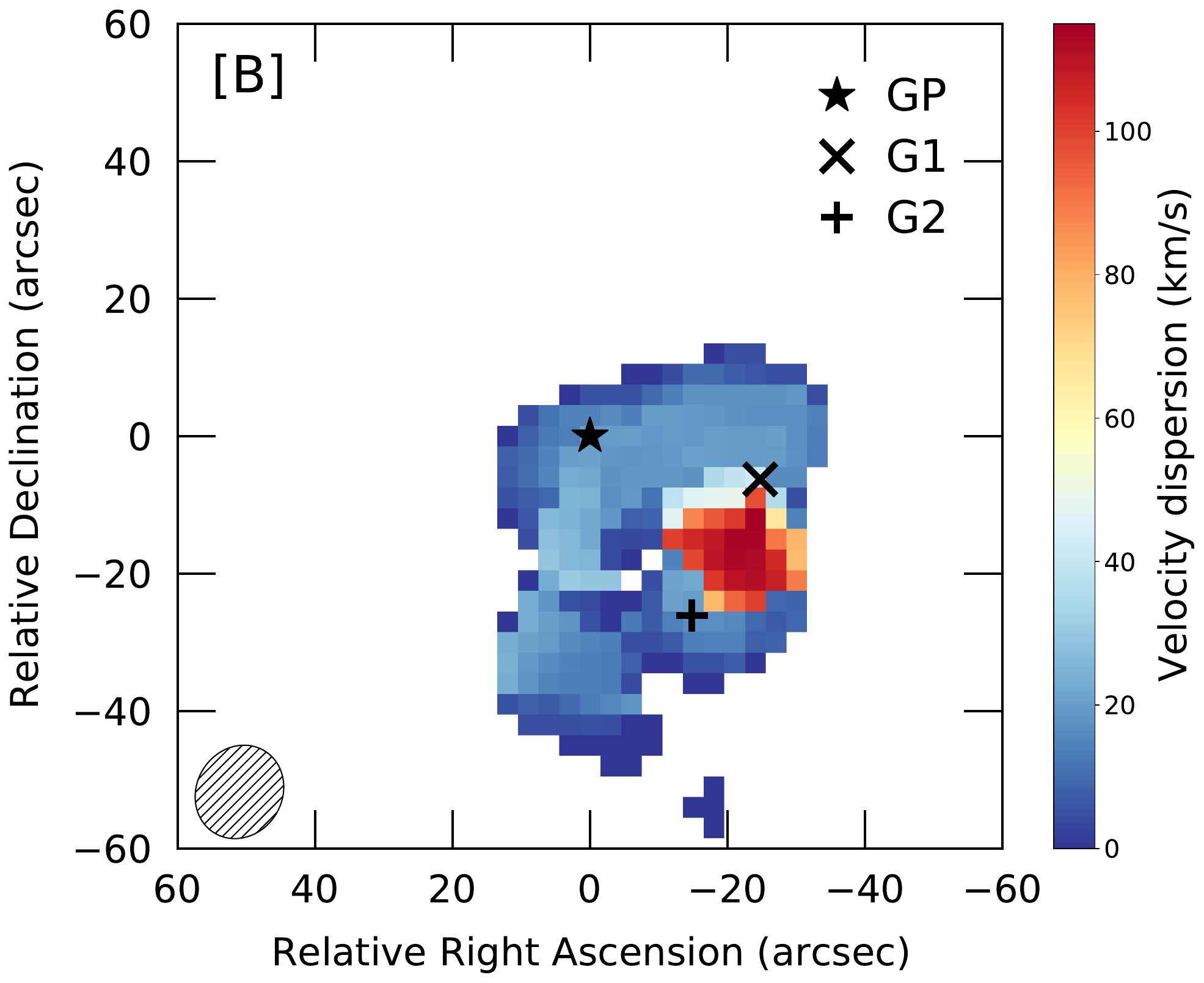}
    \caption{[A]~The \hi\ velocity field (i.e. the moment-1 image) of GP~J1148+2546 and the surrounding region, at a resolution of $14\arcsec~\times~12\arcsec$. 
    [B]~The \hi\ velocity dispersion (i.e. the moment-2 image) (in contours). The highest velocity dispersion is seen to arise in the region between galaxies G1 and G2. In both panels, the coordinates are relative to the J2000 coordinates of GP~J1148+2546, and the locations of GP~J1148+2546 and its two companions are indicated by the star, the ``X'', and the ``+'' symbols, respectively.}
    \label{fig:gp1148-mom12}
\end{figure*}

{Our VLA \hii\ image of GP~J1148+2546, from the uniformly-weighted spectral cube, is shown in Fig.~\ref{fig:gp1148-mom0}[A], in colourscale.  Fig.~\ref{fig:gp1148-mom0}[B] shows the \hii\ image in contours, overlaid on the Dark Energy Camera Legacy Survey \citep[DECaLS;][]{dey2019} g-band image (in colourscale). GP~J1148+2546 has two neighbouring galaxies in the SDSS at very close redshifts, J114825.68+254605.4 (hereafter, G1) at $z = 0.0448$, $19\arcsec$ from the GP, and J114826.34+254545.6 (hereafter, G2) at $z = 0.0456$,  $28\arcsec$ from the GP. We note that the galaxy G1 is only $\approx 17.5$~kpc from the GP in the plane of the sky. The galaxies G1 and G2 are separated by an angular distance of $\approx 21\farcs7$, approximately 1.5 times larger than the synthesized beam of the cube. The positions of the GP, G1, and G2 are indicated by a star, an ``X'', and a ``+'', respectively, in Fig.~\ref{fig:gp1148-mom0}[A], and by labels in Fig.~\ref{fig:gp1148-mom0}[B].}

The two panels of Fig.~\ref{fig:gp1148-mom0} show that the \hii\ emission arises from the region around the three galaxies, in an inverted `C' shape, with the maximum emission coming from the region between galaxies G1 and G2. {We emphasize that the angular distance between the galaxies G1 and G2 is larger than our synthesized beam, indicating that the observed peak in the \hii\ emission between G1 and G2 is not due to blending of the \hii\ peaks of the two galaxies. The \hii\ emission closest to GP~J1148+2546 is not centred on the GP, but lies to the south of the galaxy.} We also see emission $\approx 20 \arcsec$ south of the Green Pea at $> \,4\sigma$ significance. An additional galaxy, J114828.32+254542.5 (hereafter, G3) is found in the SDSS images near this \hii\ emission feature, but no spectroscopic data are available for this galaxy.

\begin{figure*}
    \centering
    \includegraphics[width=0.475\textwidth]{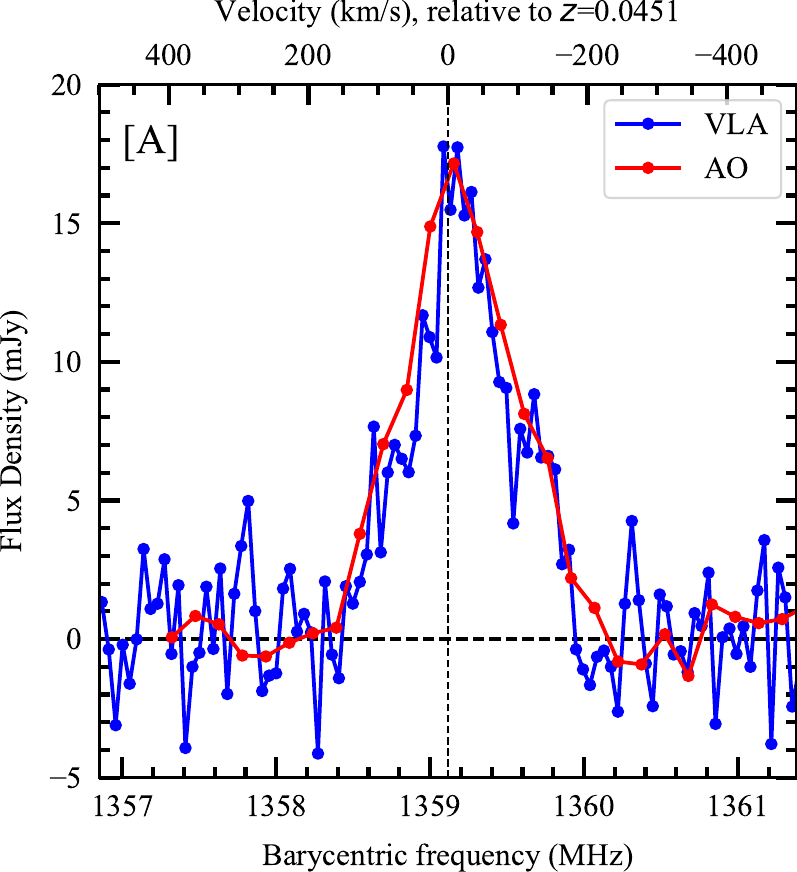}
    \includegraphics[width=0.475\textwidth]{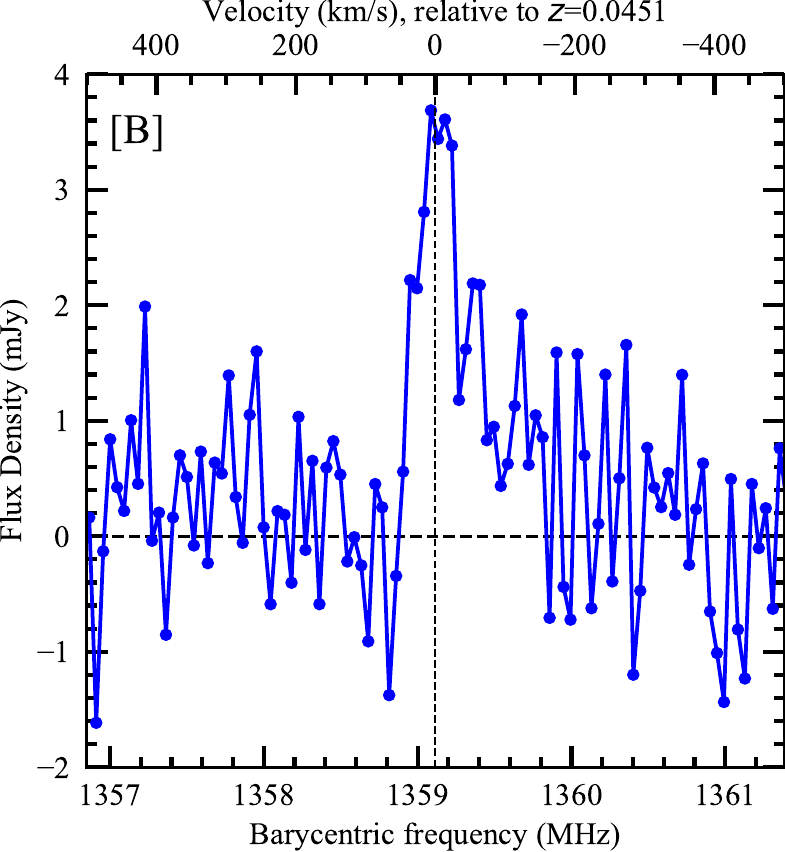}     
    \caption{[A]~The VLA \hii\ spectrum of GP~J1148+2546 (blue), overlaid on the Arecibo single-dish spectrum (red); the two are seen to be in excellent agreement. Note that the VLA spectrum includes \hii\ emission from the entire region surrounding the Green Pea, including the two companions and the extended \hii\ emission seen in Fig.~\ref{fig:gp1148-mom0}[A]. [B]~The VLA \hii\ spectrum from the region immediately around GP~J1148+2546, indicated by the circle in Fig.~\ref{fig:gp1148-mom0}[A].}
    \label{fig:gp1148-spec}
\end{figure*}

{Fig.~\ref{fig:gp1148-mom12}[A] shows the \hi\ velocity field of the gas in and around GP~J1148+2546, i.e. the first velocity moment image. It is clear that most of the extended \hi\ lies at velocities close to that of the GP, within $\approx \pm 25$~km~s$^{-1}$. The gas around the galaxy G1 is at bluer velocities, close to the systemic redshift of G1 ($z = 0.0448$), while the gas around G2 is at redder velocities, close to the systemic redshift of G2 ($z \approx 0.0456$). We note that the velocity field immediately around the Green Pea appears fairly smooth at our resolution, with no evidence for distortions that might arise due to interactions.} 

Fig.~\ref{fig:gp1148-mom12}[A] shows the \hi\ velocity dispersion of the gas, i.e. the second velocity moment image. The dispersion is seen to be fairly low, $\lesssim 20$~km~s$^{-1}$, in the gas immediately around the GP, but much larger, $\gtrsim 100$~km~s$^{-1}$, in the region between the galaxies G1 and G2. The channel maps of Fig.~\ref{fig:channelmap} show that the high observed velocity dispersion in this region is due to \hii\ emission from two different velocity ranges, $\approx -150$~\kms\ and $\approx +90$~\kms; this is also seen in Fig.~\ref{fig:pv}, a position-velocity cut across galaxies G1 and G2. 

Fig.~\ref{fig:gp1148-spec}[A] shows our VLA \hii\ spectrum of the large-scale region around GP~J1148+2546 (in blue), from the naturally-weighted cube. This was obtained by integrating over the entire region showing \hii\ emission in Fig.~\ref{fig:gp1148-mom0}. The Arecibo single-dish spectrum of \citet{kanekar2021atomic} is shown in red in the same figure. The two spectra are seen to be in fairly good agreement, implying that the VLA interferometry has not resolved out any of the Arecibo \hii\  emission.  The integrated \hii\ line flux density from the VLA spectrum is $2.734 \pm 0.023$ Jy~kms$^{-1}$, yielding an \hi\ mass of $(2.650 \pm 0.022)\times 10^{10} \ \Msun$ (note that these errors are statistical, and do not include the $\approx 10$\% uncertainty in the flux density scale). These are slightly lower than, but (given the $\approx 10$\% errors in the flux density scale) formally consistent with, the values obtained from the Arecibo \hii\ spectrum, which gave an integrated flux density of 
$3.182$ Jy~\kms\ and an \hi\ mass of $3.090 \times 10^{10} \ \Msun$ \citep[][with errors dominated by the $\approx 10$\% contribution from the flux density scale]{kanekar2021atomic}.

\begin{figure}
    \centering
    \includegraphics[width=0.475\textwidth]{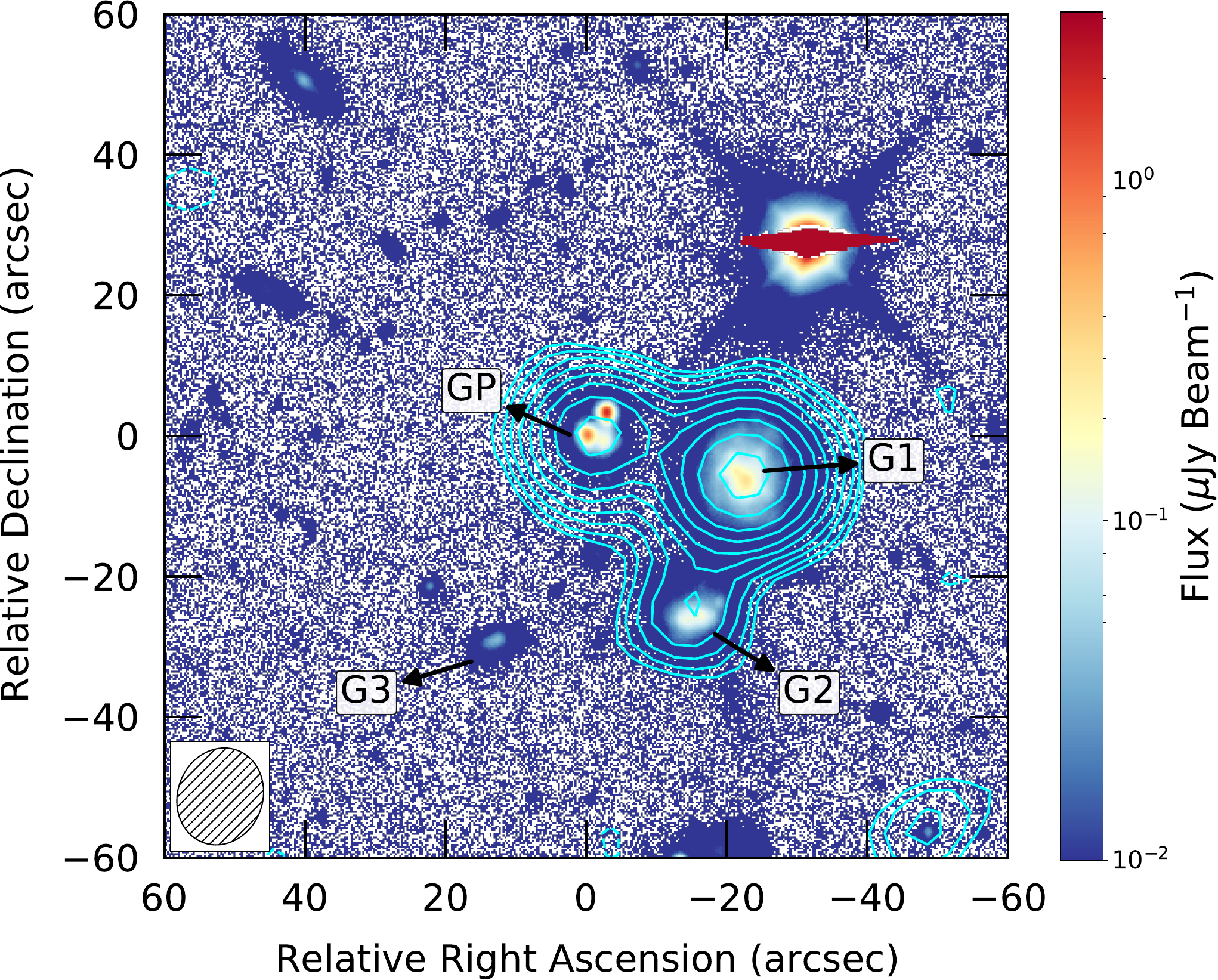}
    \caption{{The VLA 1.36~GHz continuum image of the field of GP~J1148+2546 (in contours), overlaid on the 
    DECaLS g-band image (in colour). The contour levels are [-3.0,3.0,4.2,6.0,8.4,12.0,16.8,24.0,33.6,48.0,67.2]~$\times \sigma$, where $\sigma = 36 \, \mu$Jy~Beam$^{-1}$ is the RMS noise on the VLA 1.36~GHz continuum image. The locations of GP~J1148+2546 and its companions are indicated by the labels GP, G1, and G2 (and the possible third companion by G3). The coordinates are relative to the J2000 coordinates of GP~J1148+2546.}}
    \label{fig:gp1148-cont}
\end{figure}

\begin{figure}
    \centering
    \includegraphics[width=3.5in]{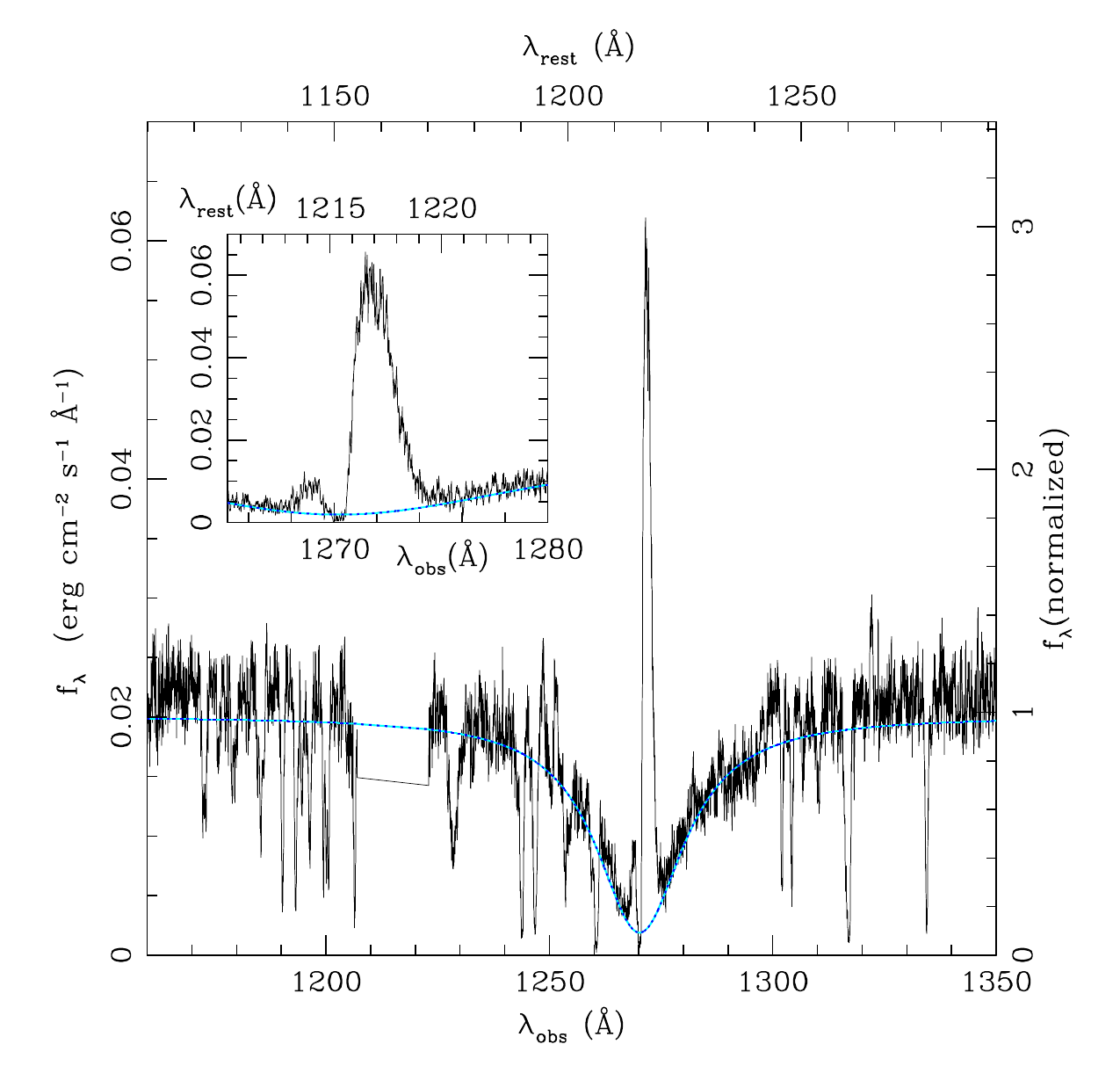}
    \caption{The HST COS spectrum of GP1148+2546 
    \citep[HST prog.~15840,][]{berg2022cos}, zooming in on the redshifted \lya\ line. The narrow \lya\ emission (also shown in the inset figure) and the wide damping wings are clearly visible. The blue/cyan dashed curve shows the Voigt profile fit to the damped \lya\ absorption.
    }
    \label{fig:gp1148-lya}
\end{figure}

{ Fig.~\ref{fig:gp1148-spec}[B] shows the VLA \hii\ spectrum of the region immediately around GP~J1148+2546 (indicated by the solid circle in Fig.~\ref{fig:gp1148-mom0}[A]), from the uniformly-weighted cube. This emission is much fainter than that in Fig.~\ref{fig:gp1148-spec}[A], indicating that GP~J1148+2546 does not contribute significantly to the total \hi\ mass of the system. We obtain a velocity-integrated \hii\ line flux density of $0.369 +/- 0.038$~Jy~\kms, for GP~J1148+2546 and its immediate surroundings. This implies an \hi\ mass of $(3.58 \pm 0.37) \times 10^9 M_\odot$, a factor of $\approx 7.4$ smaller than the total \hi\ mass of the extended \hi\ distribution measured in Fig.~\ref{fig:gp1148-spec}[A]. We further emphasize that this is formally an upper limit to the \hi\ mass of the GP as it includes contributions to the \hii\ line emission from the region around the Green Pea. The full width at half-maximum (FWHM) of the \hii\ emission is $86 \pm 12$~\kms, obtained via a single-Gaussian fit to the line profile.}

{Fig.~\ref{fig:gp1148-cont} shows our VLA 1.36~GHz continuum image of the region around GP~J1148+2546. Continuum emission is seen to be clearly detected from the Green Pea, and from both its companions. The measured 1.36~GHz flux densities of GP~J1148+2546, G1, and G2 are, respectively, $(1.286 \pm 0.055)$~mJy, $(2.890 \pm 0.064)$~mJy, and $(0.385 \pm 0.050)$~mJy. We use the relation of \citet{kennicutt12} between rest-frame 1.4~GHz luminosity and SFR to determine the radio-based SFR of GP~J1148+2546. This yields SFR$_{\rm 1.4\,GHz} = (4.00 \pm 0.17) \, M_\odot$~yr$^{-1}$. This is only slightly lower than the H$\alpha$-based SFR of $\approx 5.2 \, M_\odot$~yr$^{-1}$ estimated by \citet{jiang2019gp}. GP~J1148+2546 thus appears to be different from earlier GPs with radio continuum studies, where the radio-based SFR has been found to be significantly lower than the H$\alpha$-based SFR \citep[e.g.][]{sebastian19}.}

The HST-COS spectrum of GP~J1148+2546 is shown in Fig.~\ref{fig:gp1148-lya}, focusing on the redshifted Ly$\alpha$ line \citep[see also][for an independent analysis]{hu23}. The \lya\ emission has an equivalent width of $W_0 \approx 8$\,\AA\ and {a flux of $(1.4 \pm 0.1)\times 10^{-14}$~erg~cm$^{-2}$~s$^{-1}$}. We determined the \lya\ escape fraction by estimating the intrinsic Ly$\alpha$ line flux as 8.7 times the dust-corrected H$\alpha$ line flux, and taking the ratio of the measured Ly$\alpha$ flux to this estimate. {The inferred \lya\ escape fraction is a modest $(0.8 \pm 0.1)\%$}.  

{The \lya\ profile is double-peaked, as is common for Green Pea galaxies \citep{yang2016green,henry2015lyalpha}.   The peak separation is $2.64$\,\AA\ (observer frame), measured directly after smoothing the spectrum with a 0.45\,\AA\ boxcar filter to mitigate noise.   This corresponds to a velocity separation of $623$~\kms, placing GP~J1148+2546 on the well-established inverse relation between \lya\ escape fraction and velocity peak separation \citep{yang2017lyalpha}.   The red peak is shifted by $+236$~\kms\ from the systemic redshift, while the blue peak is shifted by $-387$~\kms.}

The prominent damped \lya\ absorption, fitted with the over-plotted Voigt profile, {yields an \hi\ column density of $(2.0 \pm 0.16) \times 10^{21}$~cm$^{-2}$} against the FUV continuum, from the fitted equivalent width and Equation~9.25 of \citet{draine11}. {The absorption trough is centered at 1270.2\AA, i.e. at a redshift of $0.04485$, which is in good agreement with the redshift of galaxy G1 ($z=0.0448$), but blueshifted by $\approx 70$~\kms\ relative to the systemic redshift of GP~J1148+2546 ($z=0.0451$).}

\begin{figure}
    \centering
	\includegraphics[width=3.4in]{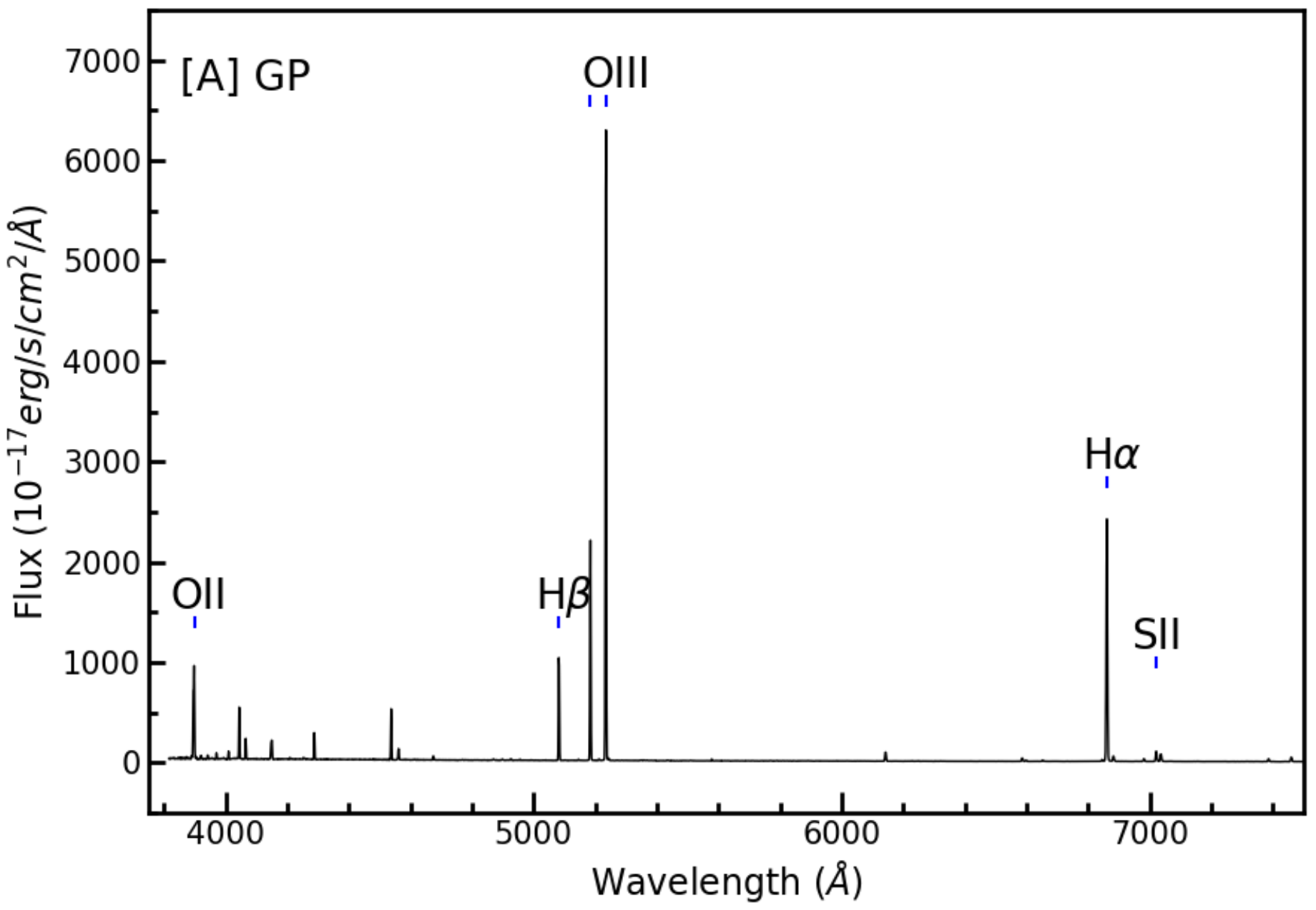}
	\includegraphics[width=3.4in]{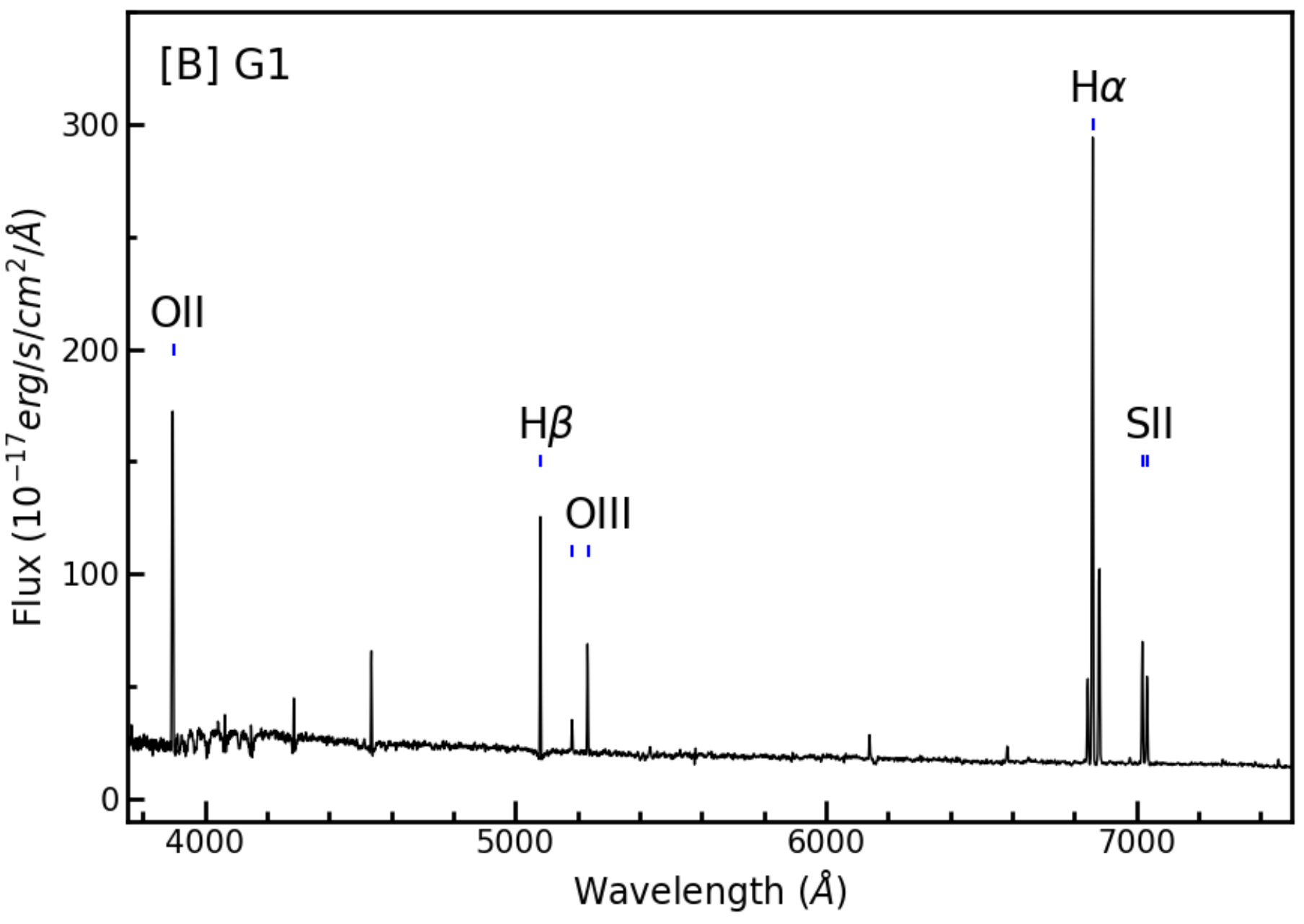}
    \caption{[A]~The SDSS spectrum of GP~J1148+2546, showing extreme emission lines, especially [O{\sc iii}]$\lambda$5007\AA. [B]~The SDSS spectrum of galaxy G1 showing strong H$\alpha$,  H$\beta$, and [O{\sc ii}]$\lambda$3727\AA\ emission lines, indicative of a normal star-forming galaxy.}
    \label{fig:gp1148-sdssspec}
\end{figure}

Fig.~\ref{fig:gp1148-sdssspec} shows the SDSS spectra of GP~J1148+2546 and the galaxy G1. The extremely strong [O{\sc iii}]$\lambda$5007\AA\ line is clearly visible in the Green Pea spectrum, with a rest equivalent width roughly 100 times higher than that in a normal star-forming galaxy (for example, the spectrum of G1 in the lower panel). We also see strong H$\alpha$ emission in the spectrum of G1, indicating significant star formation activity.

\begin{table}
\centering
\caption{{The properties of GP~J1148+2546. The rows are (1)~the redshift, (2)~the velocity-integrated \hii\ line flux density $\int S dV$, from the VLA spectrum in Fig.~\ref{fig:gp1148-spec}[B], in Jy~\kms, (3)~the \hi\ mass inferred from this spectrum, in units of $10^9 \, M_\odot$, (4)~the FWHM of the \hii\ line, W50, in \kms, (5)~the stellar mass, $M_\star$, in units of $10^8 \, M_\odot$, (6)~the \hi-to-stellar mass ratio, $f_{\rm HI} \equiv M_{\rm HI}/M_\star$, (7)~the SFR estimated from the H$\alpha$ line, in $M_\odot$~yr$^{-1}$, (8)~the SFR estimated from our measurement of the 1.36~GHz radio continuum, in $M_\odot$~yr$^{-1}$, (9)~the \hi\ depletion time, $\tau_{dep} \equiv M_{\rm HI}/{\rm SFR}$ in Gyr, (10)~the absolute blue magnitude, $\rm M_B$, (11)~the metallicity, 12+log[O/H],  (12)~O32, the ratio [O{\sc iii}]$\lambda 5007 + \lambda 4959$/[O{\sc ii}]$\lambda\lambda$3727,3729, (13)~the Ly$\alpha$ flux, $S_{Ly\alpha}$ in erg~cm$^{-2}$~s$^{-1}$, (14)~the H$\alpha$ flux, in erg~cm$^{-2}$~s$^{-1}$, (15)~the Ly$\alpha$ escape fraction $f_{\rm esc,Ly\alpha}$, (16)~the colour excess E(B-V), and (17)~the \hi\ column density towards GP~J1148+2546, measured from the Voigt profile fit to the damped Ly$\alpha$ absorption line. The redshift, stellar mass, H$\alpha$-based SFR, absolute B-magnitude, metallicity,  O32 value, and colour excess are from \citet{jiang2019gp} or \citet{senchyna17}.}
\label{tab:gp}}
\begin{tabular}{|c|c|}
\hline
\hline
$z$          & $0.0451$ \\
$\int S dV $ & $(0.369 \pm 0.038)$ Jy~\kms \\  
$M_{\rm HI}$ & $(3.58 \pm 0.37) \times 10^9 \, M_\odot$ \\
W50          & $(86 \pm 12)$~\kms \\
$M_\star$    & $5.65 \times 10^8 \, M_\odot$  \\
$f_{\rm HI}$ & $6.3 \pm 0.7$ \\
SFR$_{\rm H\alpha}$ & $5.2 \, M_\odot$~yr$^{-1}$ \\
SFR$_{1.4\,GHz}$ & $(4.00 \pm 0.17) \, M_\odot$~yr$^{-1}$ \\
$\tau_{dep}$ & $0.69$~Gyr \\
$\rm M_B$    & $-19.52$  \\ 
12+log[O/H] & $(7.96 \pm 0.04)$ \\
O32          & $5.4 \pm 0.4$ \\ 
$S_{Ly\alpha}$ & $(1.4 \pm 0.1) \times 10^{-14}$~erg~cm$^{-2}$~s$^{-1}$ \\ 
$S_{\rm H\alpha}$ & $(11.8 \pm 0.1) \times 10^{-14}$~erg~cm$^{-2}$~s$^{-1}$ \\
$f_{\rm esc,Ly\alpha}$ & $(0.8 \pm 0.1)$\%\\
E(B-V)         & $0.173$ \\
N(\hi, Ly$\alpha$) & $(2.0 \pm 0.16) \times 10^{21}$~cm$^{-2}$ \\
\hline
\hline
\end{tabular}
\vskip 0.1in
\end{table}

{The properties of GP~J1148+2546 are summarized in Table~\ref{tab:gp}. This table lists the GP redshift, the integrated \hii\ line flux density measured from the VLA spectrum of Fig.~\ref{fig:gp1148-spec}[B] for the GP and its immediate surroundings, the \hi\ mass of the GP and its immediate surroundings, the FWHM of the \hii\ line profile, the stellar mass, the \hi-to-stellar mass ratio $f_{\rm HI} \equiv M_{\rm HI}/M_\star$, the star-formation rate (SFR), the \hi\ depletion timescale, $\tau_{dep} \equiv M_{\rm HI}/{\rm SFR}$, the absolute B-band magnitude, the metallicity 12+log[O/H], the O32 value, the Ly$\alpha$ flux, and the Ly$\alpha$ escape fraction. The estimates of redshift, stellar mass, SFR, absolute B-band magnitude, metallicity, and O32 are from \citet{jiang2019gp}. We emphasize that the \hii\ properties listed in this table are for the \hii\ emission arising from GP~J1148+2546 and its immediate surroundings, and not for the total \hii\ emission from the field \citep[for the latter, see][]{kanekar2021atomic}.}

\section{Discussion}
\label{sec:discussion}

The clearly disturbed \hi\ distribution around the Green Pea indicates a merger system \citep[e.g.][]{hibbard96,sancisi99,hibbard01}. The interaction between GP~J1148+2546 and its neighbouring galaxies G1 and G2 (and possibly G3) is likely to have triggered the starburst in the Green Pea. This is reminiscent of GP~J0213+0056, where the starburst in the Green Pea is also likely to have been caused by a major merger with a companion galaxy \citep{purkayastha2022}. {However, it is curious that no signs of this interaction can be discerned in either the \hi\ velocity field or the \hi\ velocity dispersion immediately around the Green Pea, in Figs.~\ref{fig:gp1148-mom12}[A] and [B].}

Fig.~\ref{fig:gp1148-mom0}[A] shows that, unlike in the case of GP~J0213+0056 \citep{purkayastha2022}, \hii\ emission is detected at the location of GP~J1148+2546. The \hi\ column density from the VLA map at the location of  GP~J1148+2546 is $9.1 \times 10^{20}$~\cm. This is lower, by a factor of $\approx 2$, than the \hi\ column density measured from the \lya\ absorption line of Fig.~\ref{fig:gp1148-lya}; the difference is likely to arise due to the lower angular resolution of the \hii\ measurement. The high \hi\ column density is consistent with both the observed low Ly$\alpha$ escape fraction ($\approx 0.8$\%) and the relatively low value of the ratio O32~$\equiv$~[O{\sc iii}]$\lambda 5007 + \lambda 4959$/[O{\sc ii}]$\lambda$3727,3729~$\approx 5.4$ \citep{senchyna17,jiang2019gp}, which suggests low LyC and \lya\ leakage from the Green Pea \citep[e.g.][]{izotov2018low,jaskot2019}. {We emphasize that there are no direct LyC observations of GP~J1148+2546 so far, and that it is not a confirmed LyC leaker.}

\citet{kanekar2021atomic} found GP~J1148+2546 to be the strongest outlier in their GP sample (see their Fig.~3[A]) from the observed $\rm M_{HI} - M_B$ relation in the local Universe \citep{denes2014hi}. It is clear from Figs.~\ref{fig:gp1148-mom0}[A] and [B] that the estimate of \citet{kanekar2021atomic} is a significant over-estimate of the \hi\ mass of GP~J1148+2546, as the low-resolution Arecibo Telescope \hii\ spectrum includes the \hii\ emission of the neighbouring galaxies as well as the extended \hii\ emission. {We measure the \hi\ mass of GP~J1148+2546 and its immediate surroundings from the uniformly-weighted spectral cube to obtain an \hi\ mass of $(3.58 \pm 0.37) \times 10^9 \ \Msun$. For the absolute B-band magnitude of $-19.52$ of GP~J1148+2546, this is consistent with the $\rm M_{HI} - M_B$ relation of \citet{denes2014hi}. As noted in Section~\ref{sec:results}, this \hi\ mass estimate includes contributions to the \hii\ line profile from the region around the GP, and thus is formally an upper limit to the \hi\ mass of the Green Pea. The Green Pea is \hi-rich, with an \hi-to-stellar mass ratio of $\rm f_{HI} = M_{HI}/M_* \approx 6.2$.}

{The SFR of GP~J1148+2546 is $\approx 5.2 \, M_\odot$~yr$^{-1}$ \citep{jiang2019gp}. Combining this with our \hi\ mass estimate from the VLA spectrum yields an \hi\ depletion timescale of $\tau_{dep} \approx 0.69$~Gyr. This is significantly lower, by a factor $> 5$, than the typical \hi\ depletion timescales in nearby galaxies \citep[e.g., the xGASS sample; ][]{catinella18}, but similar to the median \hi\ depletion timescale of $\approx 0.6$~Gyr obtained by \citet{kanekar2021atomic} for their sample of 40~GPs at $z \approx 0.023-0.092$.}

\cite{reste2023} have recently mapped \hii\ emission from another nearby analog of EoR galaxies, Haro~11, the closest confirmed LyC-emitting galaxy (\citealp{bergvall06}; although see \citealp{grimes09}). They find a disturbed gas distribution caused by merger activity, which is likely to have facilitated the escape of LyC radiation from the galaxy. This further supports the link between merger activity and the leakage of ionizing radiation in dwarf galaxies, probably by the formation of holes or channels in the \hi\ distribution {\citep[e.g.][]{heckman11,rivera-thorsen15}. Unfortunately, the spatial resolution of our \hii\ emission images at $z \approx 0.0451$, ($\gtrsim 11$~kpc) is not sufficient to identify such putative holes in the \hi.} 

{In passing, we note that \citet{dutta24} very recently reported a tentative detection of \hii\ emission from another GP, at $z \approx 0.0326$ with a high 032 value, $\approx 15.7$. They find evidence for a spatial offset between the stellar continuum and the \hi\ distribution, suggesting a recent merger event that displaced the \hi\ from the galaxy.}

Finally, \cite{Laufman2022} conducted a search for companion galaxies in 23 GPs using the Multi~Unit Spectroscopic Explorer (MUSE) on the Very Large Telescope. They find no evidence of elevated companion fractions in GPs, compared to a control sample of main-sequence galaxies with similar stellar masses, suggesting that mergers may not be more prevalent in GPs than in normal galaxies. However, \hi\ is a much better indicator of merger activity than the dense gas traced by optical studies, making \hii\ observations essential to understand whether or not GPs have undergone a recent merger.

\section{Summary}

{We have used the VLA to map the \hii\ emission from GP~J1148+2546 at $z = 0.0451$, only the second case of \hii\ mapping of a Green Pea galaxy. We have also used archival HST-COS data to study the \lya\ spectrum of the Green Pea. The VLA \hi\ image shows that the GP is interacting with two neighbouring galaxies at very similar redshifts, the closest of which (G1) is $\approx 17.5$~kpc away. It is likely that the starburst in GP~J1148+2546 has been triggered by the interaction with its neighbours, as was earlier shown to be the case for GP~J0213+0056 at $z \approx 0.0399$. Curiously, our maps of the \hi\ velocity field and velocity dispersion do not show any signatures of the interaction at the GP location: the highest observed velocity dispersion arises between the galaxies G1 and G2.}

{Our high-resolution VLA image also shows that the bulk of the \hii\ emission detected in the Arecibo Telescope single-dish \hii\ spectrum arises from the two neighbouring galaxies and the extended \hii\ emission. We measure an \hi\ mass of $\approx (3.58 \pm 0.37) \times 10^9 \ \Msun$ for GP~J1148+2546 and its immediate surroundings, consistent with the local $\rm M_{HI} - M_B$ relation. This is a factor of $\approx 7.4$ smaller than the total \hi\ mass of the system, measured from the Arecibo \hii\ spectrum and our low-resolution VLA spectrum. We find that GP~J1148+2546 is a gas-rich galaxy, with $f_{\rm HI} \equiv M_{\rm HI}/M_\star \approx 6.3$. The Green Pea has a low \hi\ depletion timescale, $\approx 0.69$~Gyr, far lower than that of normal galaxies in the local Universe, but similar to that of the GP population at $z \lesssim 0.1$.} 

{We detect radio continuum emission from GP~J1148+2546 and its companion galaxies. We measure a 1.36~GHz flux density of $1.286 \pm 0.055$~mJy for the Green Pea, implying a radio-based SFR of $(4.00 \pm 0.17) \, M_\odot$~yr$^{-1}$. Unlike other GPs that have  been earlier studied in the radio continuum,  this is similar to the SFR of $5.2 \, M_\odot$~yr$^{-1}$ inferred from the H$\alpha$ line \citep{jiang2019gp}.}

Our HST-COS \lya\ spectrum yields a blend of narrow emission and damped absorption. The \lya\ emission has a flux of $(1.4 \pm 0.1) \times 10^{-14}$~erg~cm$^{-2}$~s$^{-1}$, with an inferred \lya\ escape fraction of $(0.8 \pm 0.1)\%$, while the damped \lya\ absorption yields an \hi\ column density of $(2.0 \pm 0.16) \times 10^{21}$~cm$^{-2}$ against the FUV continuum, a factor of $\approx 2$ higher than the \hi\ column density measured from the lower-resolution VLA \hii\ image. The modest observed \lya\ leakage is consistent with the low O32 ratio in the GP, and with the high \hi\ column density measured at the GP location in the VLA \hii\ image and from the HST-COS spectrum.

\section*{Acknowledgements}

We thank an anonymous referee for comments on an earlier draft that improved this paper. NK thanks Balpreet Kaur for much help with making the figures of this paper. The National Radio Astronomy Observatory is a facility of the National Science Foundation operated under cooperative agreement by Associated Universities, Inc. NK acknowledges support from the Department of Science and Technology via a J.~C.~Bose Fellowship (JCB/2023/000030). S.P, N.K., and S.~K. acknowledge support from the Department of Atomic Energy, under project 12-R\&D-TFR-5.02-0700. JP acknowledges funding by the Deutsche Forschungsgemeinschaft, Grant Wi 1369/31-1. This research is based on observations made with the NASA/ESA Hubble Space Telescope obtained from the Space Telescope Science Institute, which is operated by the Association of Universities for Research in Astronomy, Inc., under NASA contract NAS 5–26555.

The Dark Energy Camera Legacy Survey (DECaLS; Proposal ID 2014B-0404; PIs: David Schlegel and Arjun Dey) include data obtained at the Blanco telescope, Cerro Tololo Inter-American Observatory. This project used data obtained with the Dark Energy Camera (DECam), which was constructed by the Dark Energy Survey (DES) collaboration. Funding for the DES Projects has been provided by the U.S. Department of Energy, the U.S. National Science Foundation, the Ministry of Science and Education of Spain, the Science and Technology Facilities Council of the United Kingdom, the Higher Education Funding Council for England, the National Center for Supercomputing Applications at the University of Illinois at Urbana-Champaign, the Kavli Institute of Cosmological Physics at the University of Chicago, Center for Cosmology and Astro-Particle Physics at the Ohio State University, the Mitchell Institute for Fundamental Physics and Astronomy at Texas A\&M University, Financiadora de Estudos e Projetos, Fundacao Carlos Chagas Filho de Amparo, Financiadora de Estudos e Projetos, Fundacao Carlos Chagas Filho de Amparo a Pesquisa do Estado do Rio de Janeiro, Conselho Nacional de Desenvolvimento Cientifico e Tecnologico and the Ministerio da Ciencia, Tecnologia e Inovacao, the Deutsche Forschungsgemeinschaft and the Collaborating Institutions in the Dark Energy Survey. The Collaborating Institutions are Argonne National Laboratory, the University of California at Santa Cruz, the University of Cambridge, Centro de Investigaciones Energeticas, Medioambientales y Tecnologicas-Madrid, the University of Chicago, University College London, the DES-Brazil Consortium, the University of Edinburgh, the Eidgenossische Technische Hochschule (ETH) Zurich, Fermi National Accelerator Laboratory, the University of Illinois at Urbana-Champaign, the Institut de Ciencies de l’Espai (IEEC/CSIC), the Institut de Fisica d’Altes Energies, Lawrence Berkeley National Laboratory, the Ludwig Maximilians Universitat Munchen and the associated Excellence Cluster Universe, the University of Michigan, NSF’s NOIRLab, the University of Nottingham, the Ohio State University, the University of Pennsylvania, the University of Portsmouth, SLAC National Accelerator Laboratory, Stanford University, the University of Sussex, and Texas A\&M University.

The Legacy Surveys imaging of the DESI footprint is supported by the Director, Office of Science, Office of High Energy Physics of the U.S. Department of Energy under Contract No. DE-AC02-05CH1123, by the National Energy Research Scientific Computing Center, a DOE Office of Science User Facility under the same contract; and by the U.S. National Science Foundation, Division of Astronomical Sciences under Contract No. AST-0950945 to NOAO.

\bibliography{gp_refs}{}
\bibliographystyle{aasjournal}

\appendix
\label{sec:appendix}
\renewcommand\thefigure{A\arabic{figure}} 

\setcounter{figure}{0}
\begin{figure*}[!b]
    \centering
 \includegraphics[width=0.9\textwidth]{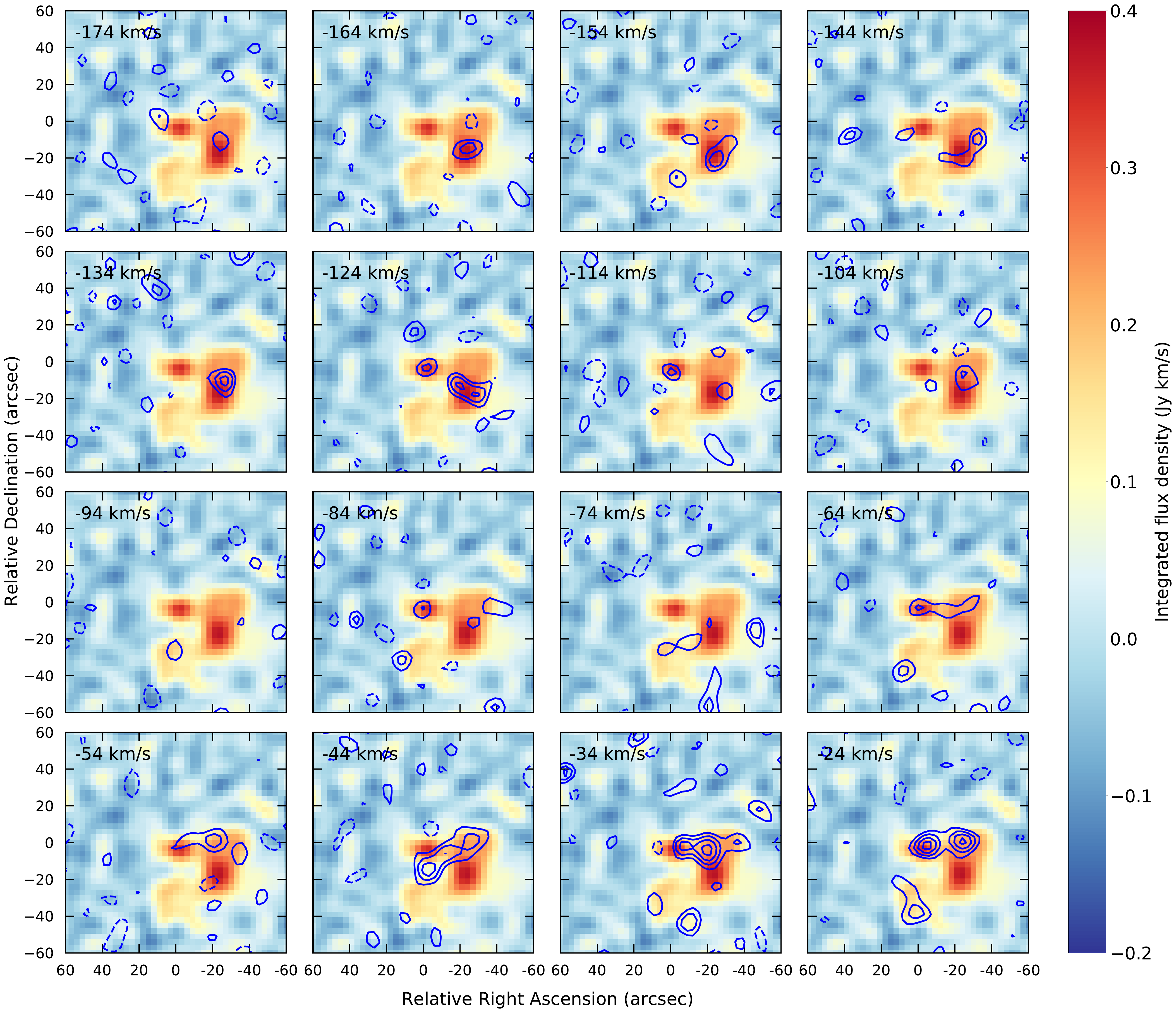}
 \caption{The contours show the channel maps at a velocity resolution of 10~\kms\ of the uniformly-weighted cube, at an angular resolution of $14\farcs0 \times 12\farcs0$. The velocity-integrated \hii\ line flux density image is shown in color scale in all panels, for comparison. The channel velocities are marked at the top left of each panel. The contours are at $(-3.0, 3.0, 4.0, 5.0, 6.0) \times \sigma$ significance, where $\sigma = 0.76$~mJy~Beam$^{-1}$ is the per-channel RMS noise.}
    \label{fig:channelmap}
\end{figure*}

\setcounter{figure}{0}

\begin{figure*}[!t]
    \centering
 \includegraphics[width=0.9\textwidth]{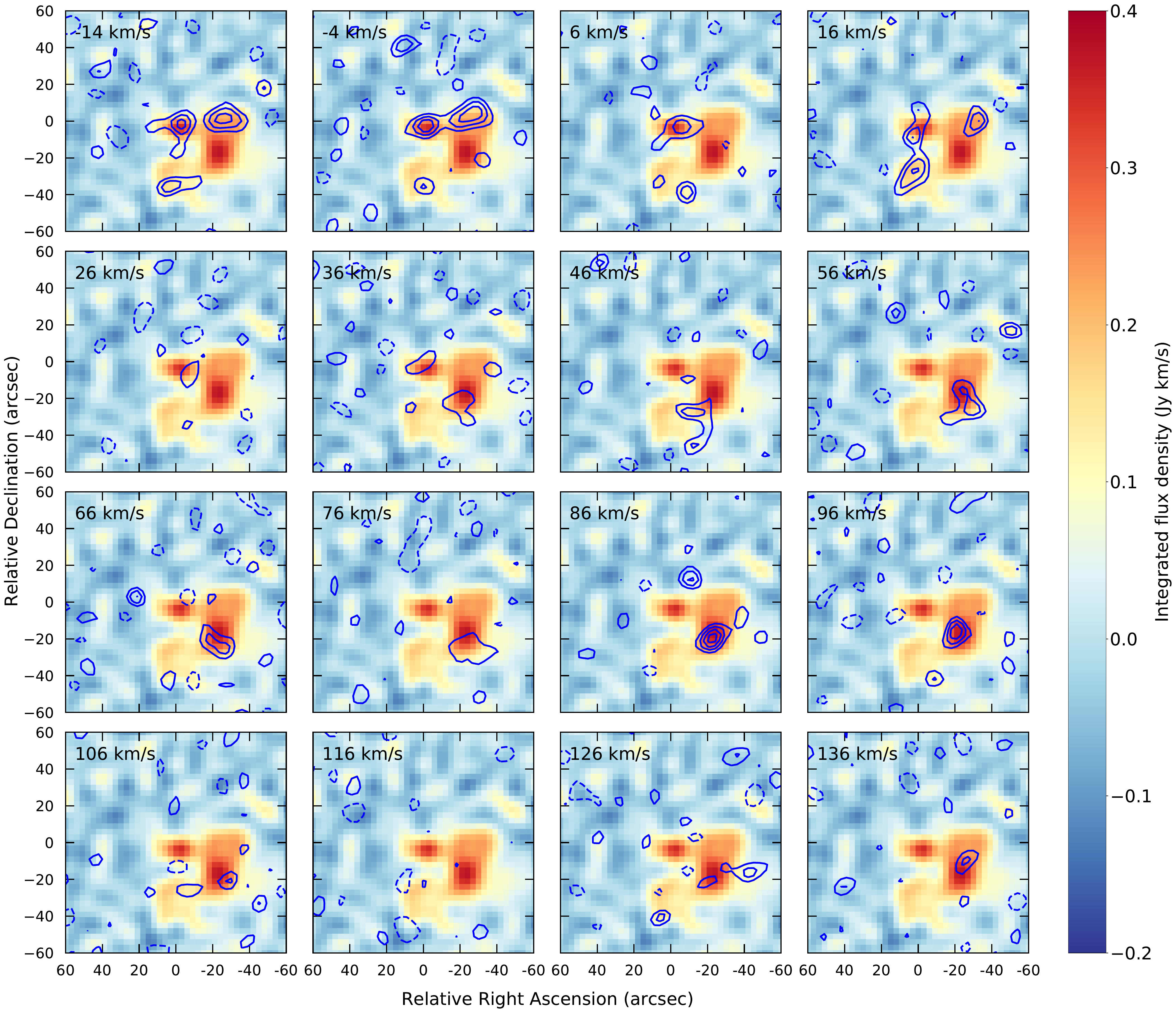}
 \caption{(contd.) The channel maps of the uniformly-weighted cube (in contours), overlaid on the velocity-integrated \hii\ line flux density image (color scale).}
    \label{fig:channelmap2}
\end{figure*}

\begin{figure*}[!t]
    \centering
 \includegraphics[width=0.49\textwidth]{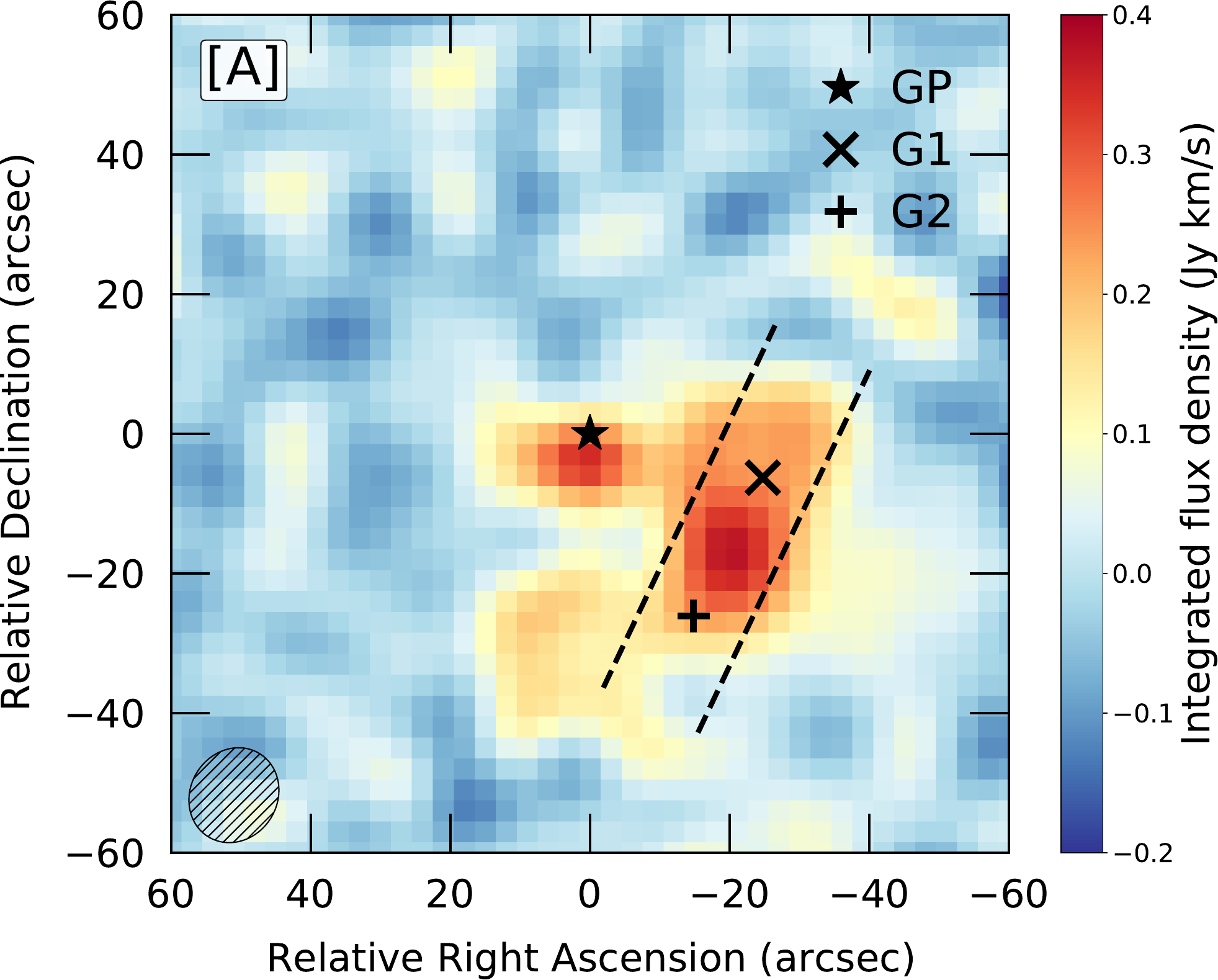}
 \includegraphics[width=0.49\textwidth]{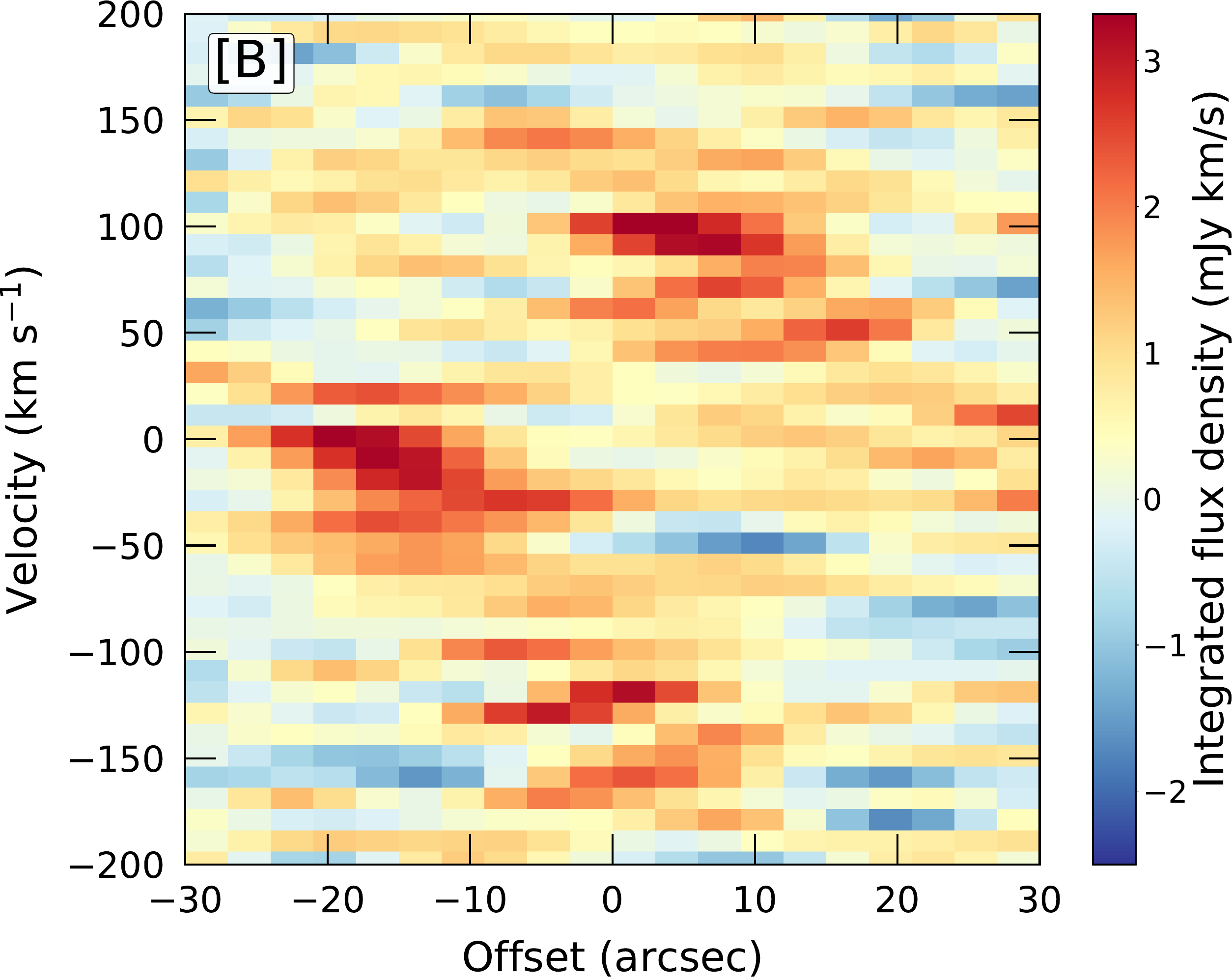}
 \caption{A position-velocity (P-V) slice across galaxies G1 and G2, and the region in between. The location of the slice is indicated by the two dashed black lines in the left panel [A], overlaid on the velocity-integrated \hii\ flux density image. The offset increases from south-east to north-west, i.e. from G2 to G1. The right panel [B] shows the resulting P-V diagram: it is clear that the central region, between galaxies G1 and G2 (at an offset of $\approx 0''$) contains emission at two velocities, $\approx -150$~\kms, and $\approx +90$~\kms.}
    \label{fig:pv}
\end{figure*}

\end{document}